\documentclass[prl,aps,twocolumn,superscriptaddress,floatfix,showpacs]{revtex4}
\usepackage{amssymb,amsmath,bm,graphicx}

\begin{document}

\title{
Low-field behavior of an $XY$ pyrochlore antiferromagnet: emergent clock anisotropies
}

\author{V. S. Maryasin}
\affiliation{Universit\'e Grenoble Alpes, INAC-PHELIQS, F-38000, Grenoble, France}
\author{M. E. Zhitomirsky}
\affiliation{CEA, INAC-PHELIQS, F-38000, Grenoble, France}
\author{R. Moessner}
\affiliation{Max-Planck-Institut f\"ur Physik komplexer Systeme, 01187 Dresden, Germany}

\date{March 4, 2016}

\begin{abstract}
Using $\rm Er_2Ti_2O_7$ as a motivation, we investigate finite-field properties of
$XY$ pyrochlore antiferromagnets. In addition to a fluctuation-induced six-fold
anisotropy present in zero  field, an external magnetic field induces a combination
of two-, three-, and six-fold clock terms as a function of its orientation providing
for a rich and controllable magnetothermodynamics. For $\rm Er_2Ti_2O_7$, we predict
a new phase transition for ${\bf H}\parallel [001]$.
Re-entrant transitions are also found for ${\bf H}\parallel [111]$.
We extend these results to the whole family the $XY$ pyrochlore antiferromagnets
and show that presence and number of low-field transitions for different
orientations can be used for locating a given material in the parameter
space of anisotropic pyrochlores. Finite-temperature classical Monte Carlo simulations
serve to confirm and illustrate these analytic predictions.
\end{abstract}
\maketitle

The Ising model is commonly used 
to describe the $Z_2$ symmetry breaking for a wide range of physical systems
including simple magnets, lattice
gases, and  neural networks \cite{Chaikin_Lubensky,Schneidman06}. Its well-known
generalizations are provided by models with $Z_N$ symmetry: Potts and clock models.
Being abundantly investigated for their own sake, these models and the related symmetry
breaking transitions \cite{Jose77} rarely appear in studies of real magnetic materials.
Yet an interesting  example of a $Z_6$ clock anisotropy was recently identified for
the ordering transition in the $XY$ pyrochlore antiferromagnet $\rm Er_2Ti_2O_7$
\cite{Champion03,Champion04,Poole07,MZ12,Savary12,Wong13,Yan13,MZ14,Ross14,McClarty14,Petit14,Javanparast15}
and in two other members of this family \cite{Li14,Dun15}.
A characteristic feature of an antiferromagnet with Ising anisotropy is the
spin-flop transition in a magnetic field applied along the easy (Ising) axis \cite{Majlis}.
Field-induced transitions in magnets with broken $Z_N$ symmetry are much less
documented. Therefore, understanding the interplay between the discrete $Z_6$ symmetry and
an external field in the $XY$ pyrochlore antiferromagnets is of significant interest from a general
perspective.

\begin{figure}[t]
\centerline{
\includegraphics[width=0.85\columnwidth]{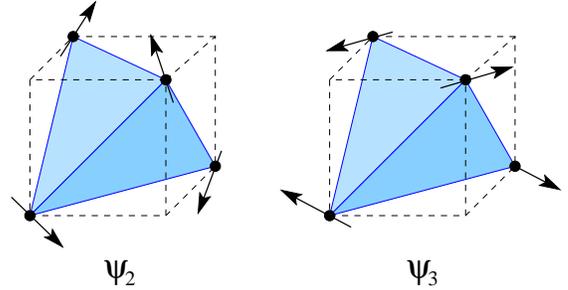}
}
\caption{(Color online)
The basis magnetic structures of the $\Gamma_5$-ordered $XY$ pyrochlores:
the noncoplanar $\psi_2$ ($m_1$) state appearing in $\rm Er_2Ti_2O_7$ in
a zero field and the coplanar $\psi_3$ ($m_2$) state.
}
\label{fig:M1M2}
\end{figure}

$\rm Er_2Ti_2O_7$ is the most studied $XY$ pyrochlore antiferromagnet. It
orders into a four-sublattice noncoplanar magnetic structure called the $\psi_2$ state
\cite{Champion03,Poole07}.
Together with a companion $\psi_3$ magnetic structure, see Fig.~\ref{fig:M1M2},
the two states transform according to the $\Gamma_5$ ($E$) representation of the tetrahedral point group $T_d$.
They remain degenerate at the mean-field level signifying an emergent $U(1)$ symmetry revealed, {\it e.g.},
in the critical behavior \cite{MZ14}.
The experimentally observed stabilization of the $\psi_2$ over the $\psi_3$ spin configuration
was attributed to an `order by disorder' effect produced by
quantum and thermal {\it fluctuations}, which generate an effective six-fold anisotropy
in the $U(1)$ manifold spanned by the $\psi_{2,3}$ states \cite{Champion03,MZ12,Savary12,Wong13}.
An alternative mechanism based on virtual crystal-field excitations
also favors the $\psi_2$ state \cite{Petit14,Mcclarty09,Rau15}.
It remains unclear at present which microscopic process dominates in
$\rm Er_2Ti_2O_7$ and,
in view of the $\psi_3$ magnetic structure found in $\rm Er_2Ge_2O_7$ \cite{Dun15},
if the selection mechanism varies across the pyrochlore family.

In this Rapid Communication, we investigate the effect of an arbitrarily oriented
field on the magnetic structure in $\Gamma_5$-ordered $XY$ pyrochlores.
The developed theory does not depend on the specific mechanism that breaks degeneracy
between $\psi_2$ and $\psi_3$ states in zero field and is applicable for both signs
of the effective six-fold anisotropy. Field-induced anisotropic terms compete with
the zero-field selection and produce orientation-dependent phase transitions.
The corresponding critical fields quantify strength of the six-fold anisotropy
in zero field, which was so far  assessed
only via the spin gap measurements \cite{Ross14,Petit14}.
Furthermore, using information about the number of phase transitions taking place for
different field orientations one can unambiguously
determine the sign of the zero-field anisotropy and position a given material
on the generalized phase diagram of anisotropic pyrochlore antiferromagnets
\cite{Wong13,Savary12b,Yan13}.

The minimal model for $\rm Er_2Ti_2O_7$ and other $XY$ pyrochlores with Kramers
magnetic ions is an effective spin-1/2 Hamiltonian
\begin{eqnarray}
\hat{\cal H} & = & \sum_{\langle ij\rangle} \bigl[ J_\perp {\bf S}^\perp_i\cdot{\bf S}^\perp_j +
J_\perp^a ({\bf S}^\perp_i\cdot \hat{\bf r}_{ij})({\bf S}^\perp_j\cdot \hat{\bf r}_{ij}) \bigr]
\nonumber \\
 & & \mbox{} - \mu_B\!\sum_i g_{\alpha\beta} H^\alpha S_i^\beta \,.
\label{H}
\end{eqnarray}
with exchange ($J_\perp$) and pseudodipolar  ($J_\perp^a$) interactions
between spin components orthogonal to the local trigonal axes $\hat{\bf z}_i$
\cite{MZ12}. Here, $\hat{\bf r}_{ij}$ denotes bond direction,
and $g_{\alpha\beta}$ is a staggered $g$-tensor with diagonal values $g_z$ and $g_\perp$.
Additional omitted terms that involve $S_i^z$ components are smaller by about
an order of magnitude \cite{MZ14,Savary12}.
We also exclude from (\ref{H}) multi-spin interactions generated by
crystal-field fluctuations \cite{Rau15}.
By fitting the low-$T$ magnetization data
for $\rm Er_2Ti_2O_7$ \cite{Bonville13}, we obtain $J_\perp=0.2$~meV, $J_\perp^a=0.3$~meV,
$g_\perp=6$, and $g_z/g_\perp \simeq 0.5$.
These values agree within 10--15\,\% with the previous estimates \cite{Savary12,Petit14}
and give good magnetization fits, see Supplemental Material for extra details
\cite{Suppl}.

Projections of a given spin configuration onto two basis states of
the $\Gamma_5$ representation denoted as $m_1$ ($\psi_2$) and $m_2$ ($\psi_3$)
form a two-component order parameter.
At the mean-field level, the bilinear spin Hamiltonian (\ref{H}) leaves a continuous degeneracy
within the $\Gamma_5$-manifold of magnetic states. The classical energy remains the same
for an arbitrary superposition of $\psi_2$ and $\psi_3$ states
characterized by $m_1\cos\varphi +m_2\sin \varphi$, thus featuring an ``accidental''
$U(1)$ symmetry. The complex combinations $m_\pm = m_1\pm im_2 = m e^{\pm i\varphi}$
transform under $T_d$ symmetry operations as
\begin{equation}
\hat{C}_3^{[111]}m_\pm = e^{\mp 2\pi i/3}m_\pm\,, \quad
\hat{\sigma}_d^{[110]} m_\pm = m_\mp\,.
\label{Sym}
\end{equation}
Allowed terms in the Landau functional correspond to invariants
constructed from  $m_1,m_2$ that are also symmetric under
time-reversal. The $\varphi$-dependent terms lift the $U(1)$ degeneracy.
In zero field, the leading degeneracy-breaking term appears at sixth order:
\begin{equation}
E_6[{\bf m}] = -\frac{a_6}{2}  (m_+^6 + m_-^6) = - A_6 \cos 6\varphi \,.
\label{E6}
\end{equation}
It is produced by a combined effect of quantum, thermal, and crystal-field fluctuations.
Specifically, 
the quantum spin-wave contribution can be
represented  by the lowest harmonics (\ref{E6}) with
\begin{equation}
A_6 = \frac{2J_\perp+J_\perp^a}{288}\,\epsilon^3 SN \,, \quad \epsilon =
\frac{J_\perp-\frac{1}{4}J_\perp^a}{J_\perp+\frac{1}{2}J_\perp^a}\,,
\label{C6}
\end{equation}
where $S=1/2$  and $N$ is the total number of sites
\cite{Maryasin14}. For $J_\perp^a/J_\perp<4$,  $A_6$ is positive selecting
the $\psi_2$ states ($\varphi_n = \pi n/3$),
whereas for $J_\perp^a/J_\perp>4$ the quantum correction  stabilizes
the $\psi_3$ states ($\varphi_n = \pi (n+\frac{1}{2})/3$).

Applying the symmetry rules (\ref{Sym}) one can also construct energy
invariants in a finite magnetic field. The lowest-order  invariant is
\begin{equation}
E_2[{\bf m},{\bf H}] =  \frac{a_2}{2}
\bigl[m_+^2\bigl(e^{2\pi i/3}H_x^2 + e^{-2\pi i/3}H_y^2 +H_z^2\bigr)
+ \textrm{c.c.} \bigr]\,.
\label{E2}
\end{equation}
An external field induces a  $2\varphi$-harmonic
in the angular-dependent part of the free energy.
For ${\bf H}\parallel [001]$ and ${\bf H}\parallel [110]$,
the expression (\ref{E2}) is further simplified to
\begin{equation}
E_2^{[001]} = A_2 H^2 \cos 2\varphi \,, \quad E_2^{[110]} = -\textstyle{\frac{1}{2}}A_2 H^2 \cos 2\varphi
\,.
\label{E22}
\end{equation}
The anisotropy has opposite signs for the two orientations,
leading to different sequences of field-induced phases and transitions.

Direct minimization  of the classical energy (\ref{H}) yields
\begin{equation}
A_2 = \frac{(g_\perp \mu_B)^2N}{8(2J_\perp + J^a_\perp)}\,.
\label{C2}
\end{equation}
Since $A_2>0$, the sole effect of magnetic field ${\bf H}\parallel [110]$ for
$\rm Er_2Ti_2O_7$ is to select two domains of the $\psi_2$ state with $\varphi=0,\pi$.
The two degenerate states smoothly evolve in an increasing field up to the transition
into a `fully polarized' state at $H=H_s$, as was observed in the neutron experiments \cite{Ruff08,Cao10}.
In contrast, for ${\bf H}\parallel [001]$, the field-induced anisotropy $E_2$ competes with the
zero-field term (\ref{E6}) producing an extra transition at $H_c = 3\sqrt{A_6/A_2}$.

\begin{figure}[t]
\centerline{
\includegraphics[width=0.75\columnwidth]{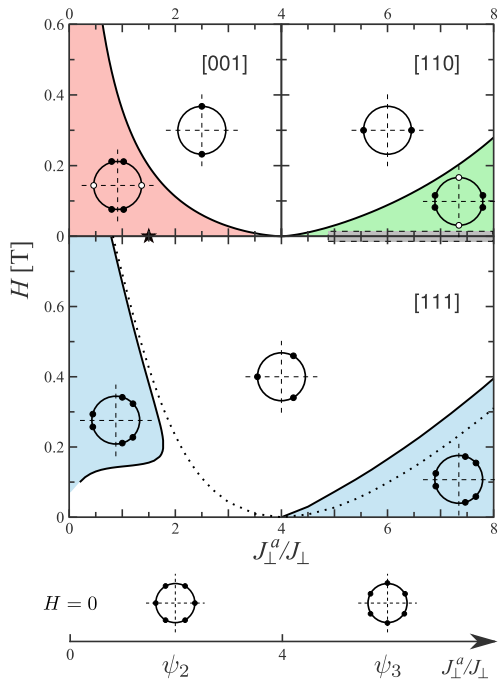}
}
\caption{Color online)
Low-field transitions in the $XY$ pyrochlore antiferromagnet
at $T=0$ as a function of $J_\perp^a/J_\perp$ ($J_\perp=0.2$~meV, $g_\perp=6$).
Top panels:  ${\bf H}\parallel[001]$ and [110]; intermediate panel:
${\bf H}\parallel[111]$.
Circles with full dots show the $U(1)$ manifold of $\Gamma_5$ states
with equilibrium values of angle $\varphi$.
Light dots denote energetically unfavorable domains. 
A star on the $J_\perp^a/J_\perp$ axis indicates the parameter ratio appropriate for $\rm Er_2Ti_2O_7$.
The shaded area for $J_\perp^a/J_\perp>4$ shows the parameter region proposed
for $\rm Er_2Ge_2O_7$ and $\rm Yb_2Ge_2O_7$. The dotted line corresponds to a single
transition exhibited by the classical model.
}
\label{fig:FD}
\end{figure}

Generally, an applied field tilts magnetic moments from the respective easy planes
and admixes other irreducible representations. These effects are, however, small
once the magnetic field is weak $H_c\ll H_s$, which is
guaranteed by small $A_6$. Accordingly, we represent
transformation of the magnetic structure in a weak field
by a dot position on a circle showing the evolution within the $U(1)$-manifold of $\Gamma_5$-states,
see Fig.~\ref{fig:FD}. In particular, below $H_c$, there are four magnetic domains
described by $4\cos^2\!2\varphi = 1 + A_2H^2/(3A_6)$.
The broken rotational symmetry is partially restored at $H=H_c$
and there remain  only two equilibrium states with
$\varphi = \pm \pi/2$.  These nearly coplanar $\psi_3$ magnetic structures
lie in the plane orthogonal to the field direction
similar to the canted spin-flop state of ordinary antiferromagnets.

We use the full expression for the field-induced anisotropy to calculate $H_c$
versus $J_\perp^a/J_\perp$ under assumption that quantum effects are dominant,
see  \cite{Suppl} for further details.
Results shown in the top-left panel of Fig.~\ref{fig:FD} were obtained with $J_\perp=0.2$~meV and $g_\perp =6$.
For $\rm Er_2Ti_2O_7$, the transition into the $\psi_3$ state takes place in a rather weak field
$H_c\approx 0.2$~T compared to $H_s=1.7$~T.
Smallness of $H_c$ reflects the strength of the order by disorder effect and
justifies the above assumptions.
By measuring $H_c$ in $\rm Er_2Ti_2O_7$,
it is possible to verify presence of other contributions
to $E_6$ produced, {\it e.g.}, by virtual crystal-field excitations \cite{Petit14,Rau15}.

For $J_\perp^a/J_\perp> 4$, the quantum anisotropy (\ref{E6}) changes sign.
Accordingly, the behavior is interchanged between the two field orientations:
continuous evolution is found for ${\bf H}\parallel[001]$, whereas an extra
transition into the $\psi_2$ state occurs for ${\bf H}\parallel[110]$, see the top-right panel
of Fig.~\ref{fig:FD}. Consequently, the identification of $\psi_{2}$ and $\psi_{3}$ magnetic states in a zero field
becomes possible by measuring the low-field transition either for ${\bf H}\parallel[110]$ or for ${\bf H}\parallel[001]$.
This may be relevant for $\rm Er_2Ge_2O_7$ and $\rm Yb_2Ge_2O_7$, which were proposed to have
the $\psi_3$ magnetic structure \cite{Li14,Dun15}.

For ${\bf H}\parallel [111]$, the second-order invariant (\ref{E2}) vanishes
in accordance with the $C_3$ rotation symmetry about the field direction, so that
selection of the phase $\varphi$ is determined by higher-order terms.
One such invariant describes the $H^2$-correction
to the  six-fold anisotropy (\ref{E6}). It comes with the
positive sign, reducing stability of $\psi_2$ states.
Explicit minimization of the classical energy (\ref{H}) yields
\begin{equation}
E_6^{[111]} = \frac{(g_\perp \mu_BH)^2N}{8(2J_\perp+J^a_\perp)}\;
\epsilon^2\cos 6\varphi = A_6'H^2\cos 6\varphi \,.
\label{E6H}
\end{equation}
Another invariant becomes important in strong fields:
\begin{equation}
E_3 [{\bf m},{\bf H}] =  \frac{a_3}{2}H_x H_y H_z (m_+^3  +  m_-^3) = A_3H^3\cos 3\varphi\,.
\label{E3}
\end{equation}
This term selects three out of six domains of the $\psi_2$ state in accordance with the
residual $C_3$ symmetry. We find numerically  $A_3>0$, hence, the three-fold anisotropy
favors $\varphi = \pm \pi/3,\pi$. Another consequence of the cubic invariant (\ref{E3}) is
changing the nature of the high-field transition into a polarized  state
from second to first order accompanied by a small magnetization jump \cite{Suppl}.

The total energy for ${\bf H}\parallel [111]$ is given by the sum of three contributions:
\begin{equation}
E(\varphi) = (A_6'H^2-A_6)\cos 6\varphi + A_3H^3\cos 3\varphi\,.
\label{Ephi}
\end{equation}
For $H\neq 0$, the stable minima of (\ref{Ephi}) correspond either to
the three $\psi_2$ states ($\varphi = \pm \pi/3,\pi$) or to the low-symmetry solutions:
\begin{equation}
\cos 3\varphi = \frac{A_3H^3}{4(A_6 - A_6'H^2)}\,.
\label{LowSym}
\end{equation}
The actual field evolution of the antiferromagnetic state
depends on sign and relative strength of the three anisotropy parameters.
For $A_6>0$ or $J_\perp^a/J_\perp<4$, the low- and the high-field states are symmetric $\psi_2$ states.
Therefore, there are either no or two consecutive phase transitions. 
For $A_6<0$, the low-symmetry solution (\ref{LowSym}) develops continuously out from the zero-field $\psi_3$ state
and upon increasing magnetic field one finds a single second-order transition into the high-field $\psi_2$-state.

We further analyzed  $E(\varphi)$ using complete analytic expression for
$A_6$ and $A_6'$ and numerically determined $A_3$ \cite{Suppl}. The obtained phase boundaries are
shown in the middle panel of  Fig.~\ref{fig:FD}. For $\rm Er_2Ti_2O_7$ with $J_\perp^a/J_\perp = 1.5$
we find two phase transitions at $H_{c1}\approx 0.15$~T and $H_{c2}\approx 0.4$~T with an intermediate
asymmetric phase. Remarkably, this material has a ratio of exchange constants close to the critical
value $(J_\perp^a/J_\perp)_c \approx 1.74$ beyond which the three $\psi_2$ states are stable in the whole range of fields.
The critical value itself depends on the strength of the $Z_6$ anisotropy and
additional contributions, {\it e.g.}, from crystal-field excitations, can reduce $(J_\perp^a/J_\perp)_c$.
In particular, observation of a double field transition in $\rm Er_2Ti_2O_7$ may be restricted
to low temperatures due to an extra contribution from thermal fluctuations.
In contrast, a single field transition for $J_\perp^a/J_\perp>4$ is a robust feature
and is present for all $T<T_c$.

In order to demonstrate presence of field-induced transitions in an unbiased way
without restrictions imposed by the analytic theory,
we also performed the classical Monte Carlo (MC) simulations of the spin Hamiltonian (\ref{H}).
The classical model lacks the $Z_6$
anisotropy  at $T=0$ and $H=0$, hence, this technique does not allow to obtain the actual
phase diagram of $\rm Er_2Ti_2O_7$. Nonetheless, an effective anisotropy is generated at finite
temperatures and the MC results are illustrative of
a generic behavior expected in real materials. In accordance with the magnetization fits,
we set $J_\perp = 1$, $J_\perp^a = 1.5$, $g_\perp = 1$, and $g_z = 0.5$. Periodic clusters
with $N=4L^3$  spins up to $L=24$ were used for simulations. The phase diagram was
reconstructed from the behavior of the total $m = \langle(m_1^2+m_2^2)^{1/2}\rangle$ and the clock order
parameters \cite{Suppl}.

\begin{figure}[t]
\centerline{
\includegraphics[width=0.6\columnwidth]{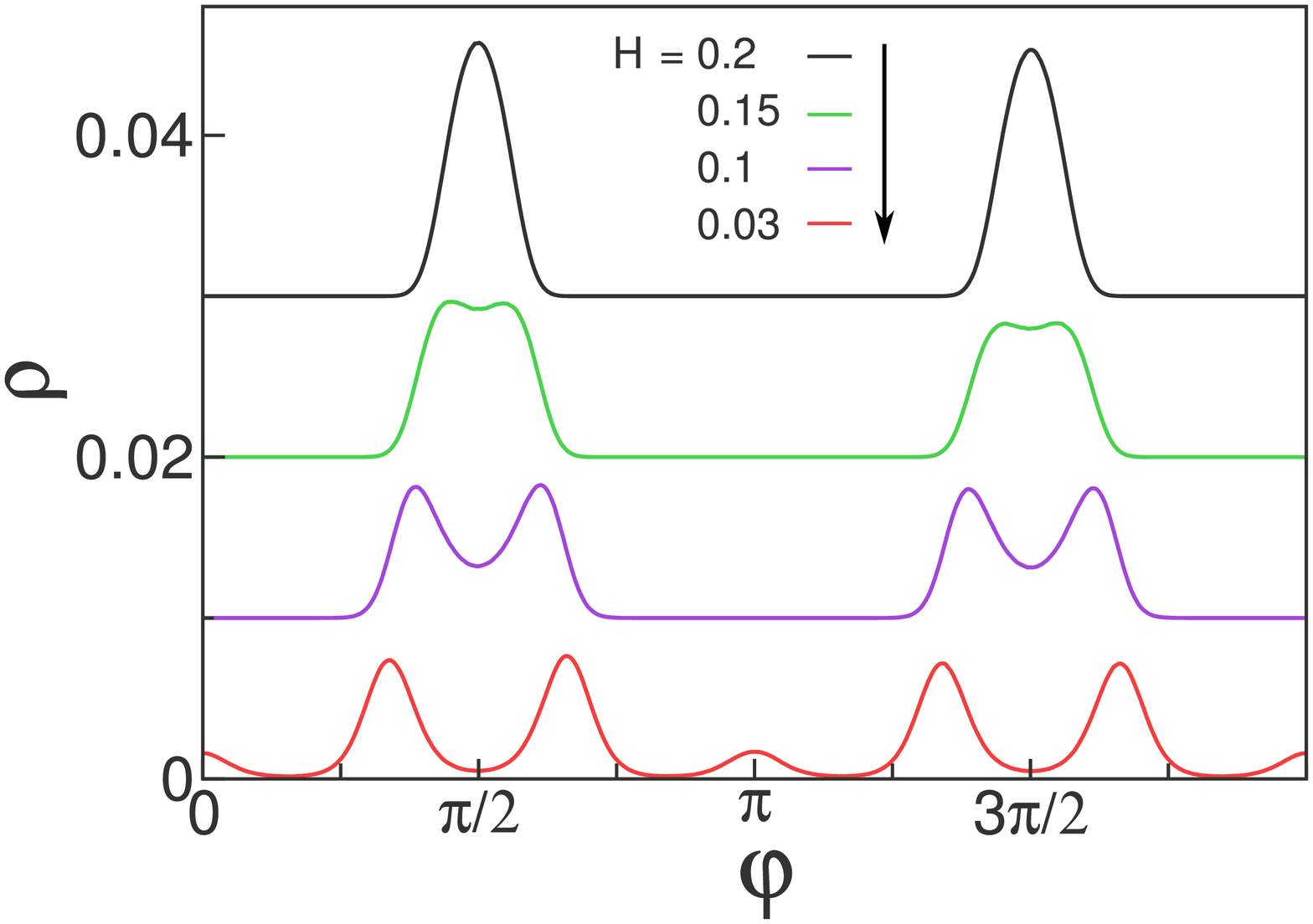} \hfill
\includegraphics[width=0.335\columnwidth]{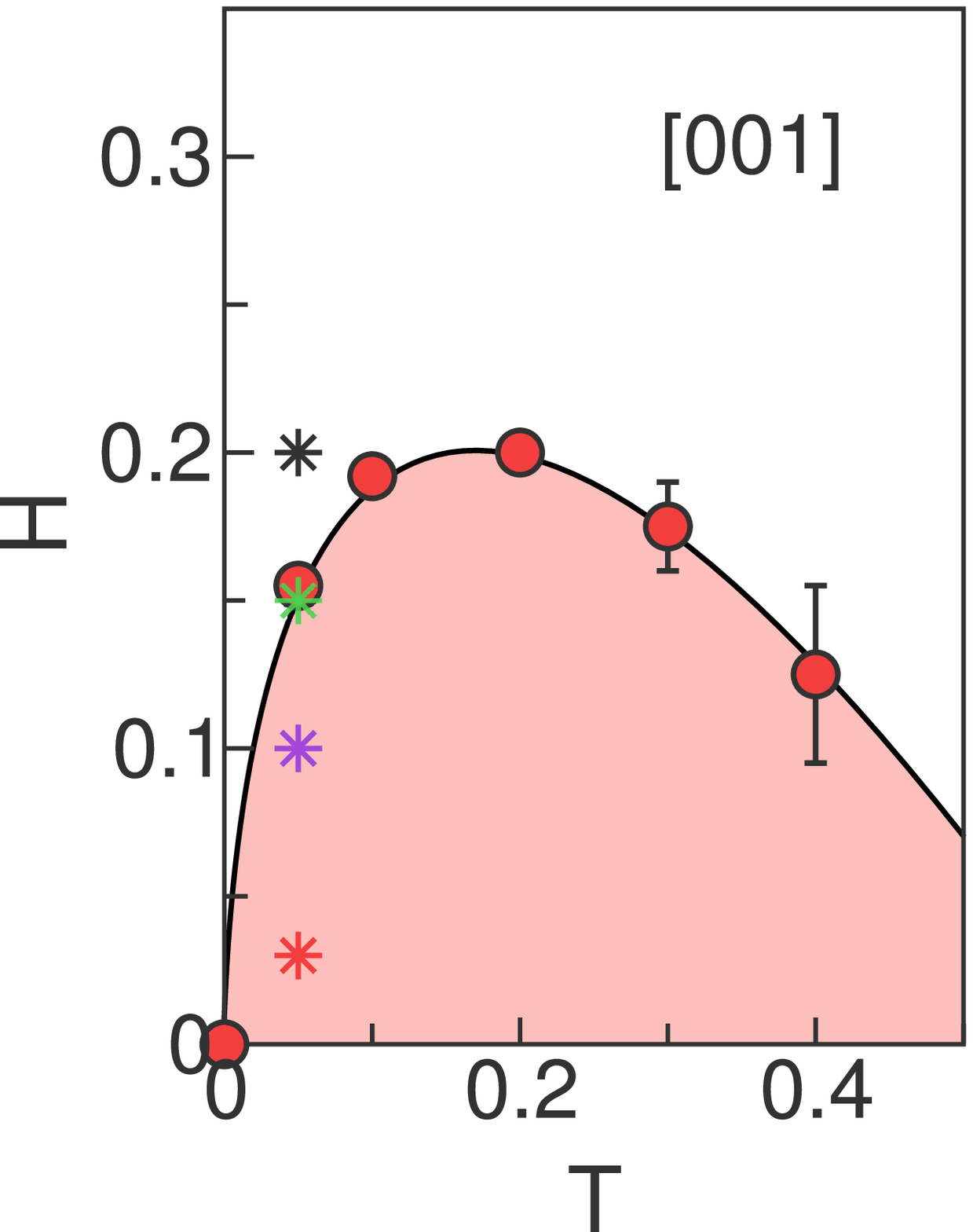}
}
\vskip 2mm
\centerline{
\includegraphics[width=0.6\columnwidth]{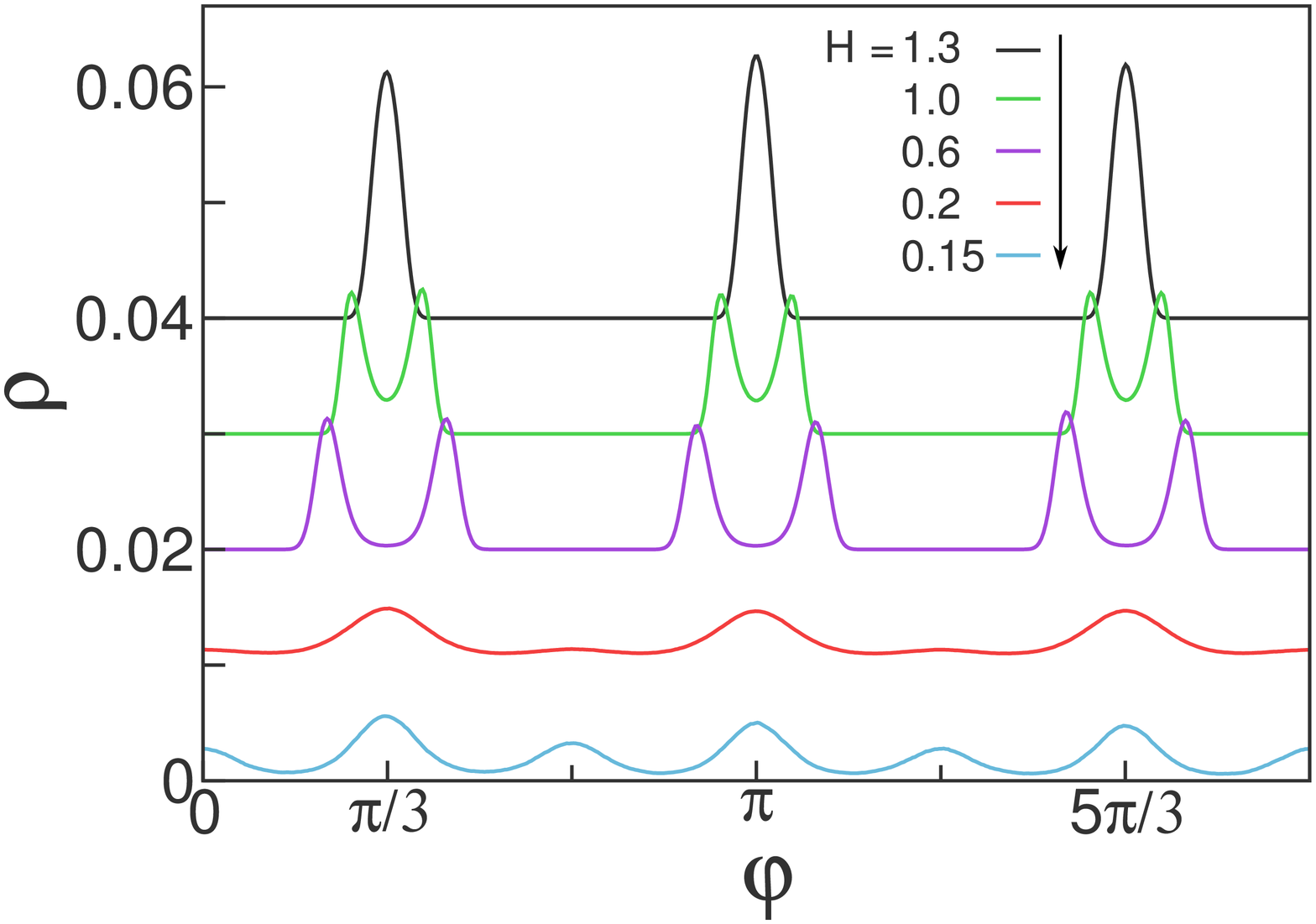} \hfill
\includegraphics[width=0.325\columnwidth]{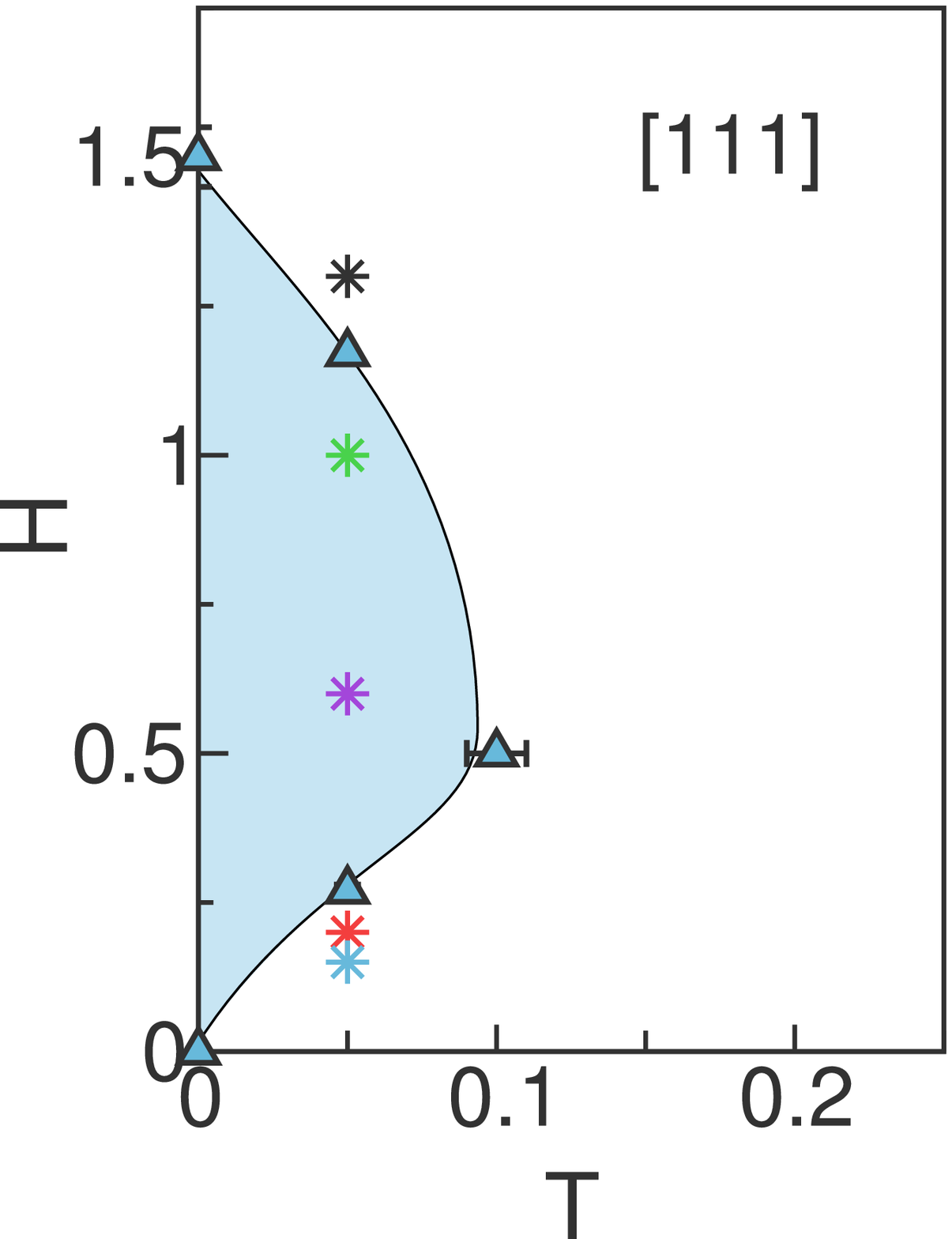}
}
\caption{(Color online) Monte Carlo results for two field orientations
$\mathbf{H} \parallel[001]$ (upper row) and $\mathbf{H} \parallel[111]$ (lower row).
Left plots show histograms for the angle $\varphi$
collected at $T=0.05$. Histograms for larger fields are progressively offset by
$\Delta\rho =0.01$.
Right plots show phase diagrams in the relevant parts of the $H$--$T$ plane.
Magnetic fields chosen for histograms are indicated by stars.
}
\label{fig:MC}
\end{figure}

The MC results the low-$T$/low-$H$ region ($T_c\approx 0.79$, $H_s(0) \approx 6.45$)
are summarized in Fig.~\ref{fig:MC}.
The transition  with a loss of the mirror symmetry for ${\bf H}\parallel [001]$ is
demonstrated by the behavior of the probability distribution function $\rho(\varphi)$
(upper left panel). At $H=0.2$ and $T=0.05$, $\rho(\varphi)$ has two sharp maxima
corresponding to the $\psi_3$ states with $\varphi=\pi/2$ and $3\pi/2$. At a lower field
$H=0.15$, each of them splits into a pair of peaks, which  move further
apart as $H$ is decreased. Using the Binder cumulant analysis we locate transition
at $H_c=0.156(2)$ \cite{Suppl}.
The temperature dependence $H_c(T)$ is shown on the upper right plot.
In the classical model, the order by disorder effect is present only at $T>0$ and the
transition field vanishes as $T\to 0$. It goes down again for
$T\to T_c$ since the six-fold anisotropy $E_6\propto m^6$ contains a higher
power of the order parameter than the field contribution $E_2\propto m^2$.

Similarly, the case of ${\bf H}\parallel [111]$ is illustrated by two lower plots in Fig.~\ref{fig:MC}.
The intermediate low-symmetry phase is evidenced by split peaks in $\rho(\varphi)$
for $H=1$ and 0.6.
The relevant part of the $H$--$T$ phase diagram is shown on the lower right plot.
The broken-symmetry state is present only for $T\leq 0.1$.  At higher temperatures, the six-fold
anisotropy generated by thermal fluctuations is sufficiently strong to restore
the $\psi_2$ states in the whole range of magnetic fields in accordance with the prior analytic
treatment.


In conclusion, using analytic symmetry arguments we demonstrated that
an external field applied to an $XY$ pyrochlore antiferromagnet
induces two-, three- or six-fold clock terms
that compete with the zero-field $Z_6$ anisotropy and
produce a remarkably rich phase diagram.
Our theory generalizes the concept of the spin-flop transition to
magnetic systems with a discrete  $Z_N$ anisotropy. Observation of such transitions is
important for determining sign and strength of the six-fold clock anisotropy
in $\rm Er_2Ti_2O_7$ and other $XY$ pyrochlores.
In particular, presence of a low-field transition for ${\bf H}\parallel [001]$
but not for ${\bf H}\parallel [110]$ unambiguously places a pyrochlore magnet into the
$J_\perp^a/J_\perp<4$ region of the parameter space ($J_{\pm\pm}>0$ in notations of \cite{Savary12})
with the $\psi_2$ magnetic structure in zero field.
The opposite behavior is expected for the $\psi_3$ ground state stabilized for  $J_\perp^a/J_\perp>4$.
These conclusions can be further corroborated by checking a number of field
transitions in the ${\bf H}\parallel [111]$ geometry (Fig.~\ref{fig:FD}).
The obtained results call for additional magnetization and polarized
neutron experiments on the $XY$ pyrochlores in a magnetic field.

\acknowledgments

We thank E. Lhotel and S. Sosin for valuable discussions and sharing their experimental data.
This work was in part supported by DFG (SFB1143).

\onecolumngrid

\mbox{}

\clearpage

\begin{center}
\large\bf
SUPPLEMENTAL MATERIAL
\end{center}

\setcounter{equation}{0}

\section{I.\ Extracting model parameters from magnetization curves}
\label{Sec:Fit}

In $\rm Er_2Ti_2O_7$,  the lowest Kramers doublet of Er$^{3+}$ ions selected by the crystal field
is separated from the next two levels by gaps of 6.38 and 7.39~meV \cite{Champion03s}.
Accordingly, the minimal spin model for this material applicable at low temperatures and
weak magnetic fields can be formulated in terms of the pseudo spin-1/2 operators acting in
the subspace of ground-state Kramers doublets. This assumption is corroborated by the recent
heat-capacity measurements \cite{Niven14s,Dalmas12s}, which find
a pronounced $R\ln 2$ plateau in the temperature dependence of the magnetic entropy $S_{\rm mag}(T)$ below 10~K.
The effective nearest-neighbor Hamiltonian is a bilinear form of these spin-1/2 operators \cite{Zhitomirsky12s}:
\begin{equation}
\hat{\cal H} =  \sum_{\langle ij\rangle}\bigl[
J_\perp {\bf S}^\perp_i\cdot{\bf S}^\perp_j + J_\perp^a
({\bf S}^\perp_i\cdot \hat{\bf r}_{ij})({\bf S}^\perp_j\cdot \hat{\bf r}_{ij})
\bigr] - \sum_i   g_{\alpha\beta}\mu_B\,  H^\alpha S_i^\beta\,.
\tag{S1}
\label{Hmzh}
\end{equation}
Here ${\bf S}^\perp_i$ are spin components perpendicular to the
local trigonal axis $\hat{\bf z}_i$, $\hat{\bf r}_{ij}$ is a unit vector in the bond direction,
and $g_{\alpha\beta}$ is a staggered $g$-tensor with the uniaxial symmetry:
\begin{equation}
g_{\alpha\beta}=g_z \hat{z}_i^\alpha \hat{z}_i^\beta + g_\perp \bigl(\delta_{\alpha\beta}- \hat{z}_i^\alpha \hat{z}_i^\beta\bigr) \,.
\tag{S2}
\end{equation}
In accordance with the $XY$ nature of the Er$^{3+}$ magnetic moments, the effective Hamiltonian (\ref{Hmzh}) contains
only planar components of spins. The omitted terms that include the $S_i^z$ components are smaller by about an order of magnitude \cite{Zhitomirsky12s,Savary12s}.

\begin{figure}[b]
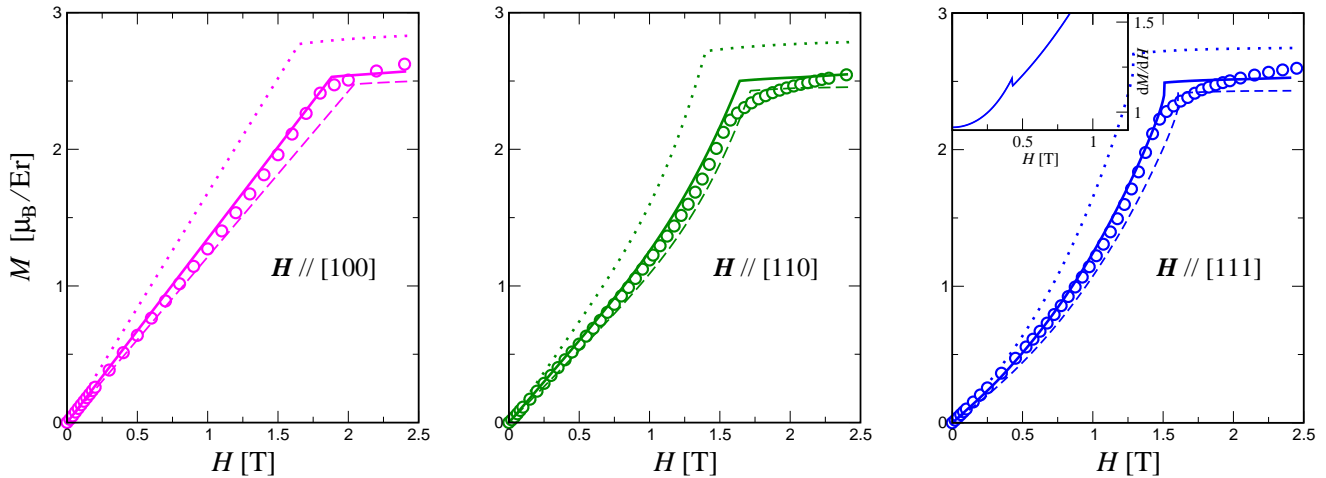

\centerline{
\includegraphics[height=0.35\columnwidth]{magnetization100}
\hskip 2mm
\includegraphics[height=0.35\columnwidth]{magnetization110}
\hskip 2mm
\includegraphics[height=0.35\columnwidth]{magnetization111}
}
\caption{Magnetization of $\rm Er_2Ti_2O_7$ for three field orientations.
Data points are the experimental results of Bonville
{\it et al.}~\cite{Bonville13s} taken at $T=175$~mK ([100]) and 130~mK ([110] and [111]).
The zero-temperature magnetization curves for the effective spin-1/2 Hamiltonian with the parameters given by Eq.~(\ref{HParam})
are shown by the full lines. The dashed lines correspond to theoretical curves obtained with the parameters
of Savary~{\it et al.}~\cite{Savary12s}. The dotted lines are drawn for the set of coupling constants
by Petit {\it et al}.~\cite{Petit14s,Cao09s}.
The inset on the right panel displays the field derivative
$dM/dH$ in the low field region.
}
\label{fig:Mfit}
\end{figure}

To fix parameters of the spin Hamiltonian (\ref{Hmzh}) we have fitted
the low-temperature magnetization of $\rm Er_2Ti_2O_7$ reported by Bonville {\it et al.}~\cite{Bonville13s}.
In all field orientations the measured  $M(H)$ curves exhibit weak almost linear growth
in a polarized paramagnetic state above the critical field $H_s\simeq 1.5$--1.7~T.
We have subtracted this isotropic Van Vleck susceptibility $\chi_{V}/\textrm{Er}\approx 0.2\mu_B/\textrm{T}$
from the original experimental data. Magnetization in a given
field has been computed for a classical spin configuration obtained by minimizing numerically the energy (\ref{Hmzh})
within the subspace of the four-sublattice magnetic structures.
The experimental data together with our theoretical fits are shown in Fig.~\ref{fig:Mfit} for three field orientations.
The microscopic parameters deduced from these fits are
\begin{equation}
J_\perp = 0.2~\textrm{meV}\,, \quad J_\perp^a = 0.3~\textrm{meV}\,, \quad g_z = 3.0\,, \quad g_\perp = 6.0\,.
\tag{S3}
\label{HParam}
\end{equation}

Previously, the exchange parameters of $\rm Er_2Ti_2O_7$ were obtained by Savary and coworkers
\cite{Savary12s} from the high-field INS measurements. They found
$J_\perp = 0.21(2)$~meV, $J^a_\perp = 0.35(5)$~meV, $g_z = 2.45\pm0.23$ and $g_\perp = 5.97\pm 0.08$
with two additional coupling constants $J_{zz} = -0.025(2)$~meV, $J_{z\perp} = 0.03(5)$~meV omitted
in the minimal spin model (\ref{Hmzh}). The corresponding magnetization curves are shown by dashed lines.
An independent set of the coupling constants was derived by Petit {\it et al}.~\cite{Petit14s} from
the zero-field INS experiments:
$J_\perp = 0.206$~meV, $J^a_\perp = 0.308$~meV and also $J_{zz} = -0.022$~meV, $J_{z\perp} = 0.052$~meV.
Using the $g$-tensor values obtained earlier by the same group  \cite{Cao09s},
$g_z = 2.6$ and $g_\perp = 6.8$, we plot the resulting magnetization curves by dotted lines
in Fig.~\ref{fig:Mfit}.
Overall, the three sets of microscopic parameters match each other within 10--15\%.
Nevertheless,  our set (\ref{HParam}) demonstrates better agreement with the magnetization data 
justifying at the same time the use of the minimal spin model (\ref{Hmzh}).

The theoretical magnetization curve for ${\bf H}\parallel [111]$ in Fig.~\ref{fig:Mfit} exhibits 
a small but clear jump at the saturation field $H_s^{[111]}\approx 1.5$~T. This weak first-order 
transition into the polarized paramagnetic state is a direct consequence of the cubic invariant in 
the Landau energy functional for the $\Gamma_5$ order parameter given by Eq.~(9) in the main text. 
The invariant vanishes for the two other field orientations leaving in those cases continuous second-order 
transitions at $H_s$.

The inset of the right panel of Fig.~\ref{fig:Mfit} shows the field derivative $dM/dH$ of the theoretical 
magnetization curve for ${\bf H}\parallel [111]$. A small jump in the derivative at $H_c\approx 0.4$~T 
indicates a second-order phase transition. We identify the anomaly with a transition between a low-symmetry 
state, Eq.~(11) of the main text, and the $\psi_2$ states with $\varphi =\pi$ and $\pm \pi/3$ at high fields. 
Presence of such a transition can be easily understood on the basis of Eq.~(10) in the main text with
$A_6\equiv 0$, see also Sec.~III below. This transition is present for all ratios of $J_\perp^a/J_\perp$ and 
is shown in Fig.~2 of the main text by a dotted line.

\section{II.\ Energy Correction in a weak magnetic field}

Here we outline the analytic derivation and give complete expressions
for the state-dependent energy corrections induced by a weak magnetic field.
We adopt the following convention for the positions of magnetic atoms
 in the unit cell of a pyrochlore lattice
\begin{align}
\mathbf{r}_1 = (0,0,0)\,, \quad
\mathbf{r}_2 = (\tfrac{1}{4},\tfrac{1}{4},0)\,, \quad
\mathbf{r}_3 = (0,\tfrac{1}{4},\tfrac{1}{4})\,, \quad
\mathbf{r}_4 = (\tfrac{1}{4},0,\tfrac{1}{4})\,. \quad
\tag{S4}
\end{align}
The local coordinate frame for each site is defined by the set of basis vectors
\begin{align}
&\mathbf{\hat{x}}_1 = \frac{1}{\sqrt{6}}(1,1,-2)\,,  \  && \mathbf{\hat{x}}_2 = \frac{1}{\sqrt{6}}(-1,-1,-2)\,, \  && \mathbf{\hat{x}}_3 = \frac{1}{\sqrt{6}}(1,-1,2)\,, \  && \mathbf{\hat{x}}_4 = \frac{1}{\sqrt{6}}(-1,1,2)\,, \nonumber  \\
&\mathbf{\hat{y}}_1 = \tfrac{1}{\sqrt{2}}(-1,1,0)\,, \  && \mathbf{\hat{y}}_2 = \frac{1}{\sqrt{2}}(1,-1,0)\,, \  &&
\mathbf{\hat{y}}_3 = \frac{1}{\sqrt{2}}(-1,-1,0)\,,  \  && \mathbf{\hat{y}}_4 = \frac{1}{\sqrt{2}}(1,1,0)\,, 
\tag{S5} \label{Axes}
\\
&\hat{\bf z}_1= \frac{1}{\sqrt{3}}(1,1,1)\,,  \  &&
\hat{\bf z}_2 = \frac{1}{\sqrt{3}}(-1,-1,1)\,,  \  &&
\hat{\bf z}_3 = \frac{1}{\sqrt{3}}(1,-1,-1)\,,  \  &&
\hat{\bf z}_4 = \frac{1}{\sqrt{3}}(-1,1,-1)\,.  \ \nonumber
\end{align}
The $\hat{\bf x}_i$ and $\hat{\bf y}_i$ axes coincide with spin directions in the $\psi_2$ and $\psi_3$ magnetic structures, respectively,
see Fig.~1 of the main text. The degenerate $E$ ($\Gamma_5$) manifold of ground-state spin configurations can be
parameterized by a single angle: $\mathbf{S}_i = \hat{\bf x}_i\cos\varphi + \hat{\bf y}_i\sin\varphi$.
Expansion of the Hamiltonian (\ref{Hmzh}) in small deviations from a classical ground state was performed in
Ref.~\cite{Maryasin14s}.  Keeping only terms that are quadratic in  in-plane, $\mathsf{S}^y_i$, and
out-of-plane, $\mathsf{S}^z_i$, spin components we rewrite (\ref{Hmzh}) as
\begin{equation}
\hat{\cal H}_2 = \frac{h}{2S} \sum_i \left[ \mathsf{S}_i^z{}^2 + \mathsf{S}_i^y{}^2 \right] -
2\sum_{\langle ij \rangle}M_{ij}\,\mathsf{S}_i^y\,\mathsf{S}_j^y \,; \qquad i,j = 1 \ldots 4\,,
\tag{S6}
\label{Ham}
\end{equation}
where $h = (2J_\perp + J_\perp^a)S$ is a site-independent amplitude of the local magnetic field
and $M_{ij}$ is a set of bond-dependent coupling constants
\begin{align}
& M_{ij} = \frac{1}{6}\,(2J_\perp + J_{\perp}^a) + \frac{1}{6}\,(4J_\perp - J_{\perp}^a)
\cos (2\varphi + \gamma_{ij})\,, \nonumber \\
&\gamma_{12} = \gamma_{34} = 0\,, \qquad \gamma_{13} = \gamma_{24} = \frac{2\pi}{3}\,, \qquad
\gamma_{14} = \gamma_{23} = -\frac{2\pi}{3}\,.
\tag{S7}
\label{Phases}
\end{align}

An external magnetic field adds linear terms to $\hat{\cal H}_2$, which distort the magnetic structure.
In the rotated local frame the Zeeman energy becomes
\begin{equation}
\hat{\cal H}_Z = -\sum_i \left[ g_z \mu_B H_i^z \mathsf{S}_i^z + g_\perp \mu_B H_i^\perp\mathsf{S}_i^y\sin(\varphi - \phi_i) + O(\mathsf{S}_i^2)\right],
\tag{S8}
\label{Zeeman}
\end{equation}
where $\phi_i$ is an angle between the in-plane component of the applied field and
the $\hat{\bf x}_i$ spin axis. Explicitly, the in-plane field components
$H^\perp_i$ and the polar angles $\phi_i$ are given for the three field orientations by
\begin{align}
H &\ \parallel  [001] & & & H &\ \parallel  [110] & & & H &\ \parallel  [111] & &  \nonumber \\
H_1^\perp & = \sqrt{\tfrac{2}{3}}H; & \phi_1 & = \pi;\qquad\qquad   & H_1^\perp & = \tfrac{1}{\sqrt{3}}H; & \phi_1 & = 0;\qquad\qquad  & H_1^\perp & = 0; & \nonumber \\
H_2^\perp & = \sqrt{\tfrac{2}{3}}H; & \phi_2 & = \pi;\qquad\qquad   & H_2^\perp & = \tfrac{1}{\sqrt{3}}H; & \phi_2 & = \pi;\qquad\qquad  & H_2^\perp & = \tfrac{\sqrt{8}}{3}H; & \phi_2 & = \pi; 
\tag{S9}
\label{FieldDir} \\
H_3^\perp & = \sqrt{\tfrac{2}{3}}H; & \phi_3 & = 0;\qquad\qquad   & H_3^\perp & = H; & \phi_3 & = -\pi/2;\qquad\qquad  & H_3^\perp & = \tfrac{\sqrt{8}}{3}H; & \phi_3 & = -\pi/3; \nonumber\\
H_4^\perp & = \sqrt{\tfrac{2}{3}}H; & \phi_4 & = 0;\qquad\qquad   & H_4^\perp & = H; & \phi_4 & = \pi/2;\qquad\qquad  & H_4^\perp & = \tfrac{\sqrt{8}}{3}H; & \phi_4 & = \pi/3. \nonumber
\end{align}

Since the terms containing $\mathsf{S}^z_i$ in Eqs.~(\ref{Ham}) and (\ref{Zeeman}) are $\varphi$-independent,
the out-of-plane deviations do not affect the ground state degeneracy at quadratic order in $H$.
To minimize over the in-plane fluctuations $\mathsf{S}^y_i$ we first diagonalize of the quadratic form (\ref{Ham})
with the help of a suitable  orthogonal transformation:
\begin{equation}
T = \frac{1}{2}
\begin{pmatrix}
-1 &       -1 & \;\;\,1 & 1 \\
-1 & \;\;\, 1 &      -1 & 1 \\
\;\;\,1 &  -1 &      -1 & 1 \\
\;\;\,1 &\;\;\,1 &\;\;\,1 & 1
\end{pmatrix}.
\tag{S10}
\end{equation}
The eigenvalues $\Lambda = \rm{diag} \lbrace \lambda_1; \lambda_2; \lambda_3; \lambda_4 \rbrace$
calculated as $\Lambda = T^{-1}MT$ are given by
\begin{equation}
\lambda_n = \frac{2}{3}(2J_\perp + J_{\perp}^a) - \frac{1}{3}(4J_\perp - J_{\perp}^a)\cos (2\varphi + \gamma_{1\,n+1})\,,
\ \  n=1\! - \!3\,,  \qquad \lambda_4 = 0\,.
\tag{S11}
\label{lambda}
\end{equation}
The zero eigenmode $\lambda_4$ corresponds to motion of spins inside the degenerate ground-state manifold
and does not contribute to the degeneracy lifting.
The energy correction is, then, obtained by direct minimization of the diagonal quadratic form and the linear terms:
\begin{equation}
\Delta E = - \frac{(g_\perp\mu_B)^2}{4} \frac{N}{4} \sum_{n=1}^3 \frac{\Bigl(\sum_{k=1}^4 H_k^\perp \sin(\varphi - \phi_k) t_{kn}
\Bigr)^2}{\lambda_n} \,.
\tag{S12}\label{DEZ}
\end{equation}
Here $t_{kn}$ are elements of the transformation matrix $T$ and the normalization factor $N/4$ extends
the result to the entire lattice with $N$ spins.
Full expressions for the $\varphi$-dependent energy terms quadratic in a weak magnetic field $H$
are obtained by substituting  $H^\perp_k$ and $\phi_k$ from Eq.~(\ref{FieldDir}).

For a magnetic field applied along the $[001]$ axis, the field-induced energy correction is given by
\begin{equation}
\Delta E^{[001]} = \frac{(g_\perp \mu_B H)^2 N}{8(2J_\perp + J_\perp^a)} \  \frac{\cos 2\varphi-1}{1-\epsilon\cos 2\varphi} \,,
\tag{S13}\label{DEZ001}
\end{equation}
where $\epsilon$ is a dimensionless parameter $\epsilon = (J_\perp-J_\perp^a/4)/(J_\perp+J_\perp^a/2)$.
In the parameter region spanned by positive exchange constants $J_\perp$ and $J_\perp^a$, the variations of $\epsilon$ are
restricted to $-0.5 < \epsilon < 1$. In particular, for $\rm Er_2Ti_2O_7$  we obtain
$\epsilon\approx 0.36$.
Thus, for all relevant $\epsilon$, the energy correction remains finite and selects states with $\varphi=\pm\pi/2$.
The expression for $\Delta E^{[001]}$  presented in the main text corresponds to $\epsilon =0$.
To achieve a better accuracy one may use the full expression (\ref{DEZ001}).

For $\mathbf{H} \parallel [110]$ the field-induced anisotropy has a  more complex expression
\begin{equation}
\Delta E^{[110]} = - \frac{(g_\perp \mu_B H)^2 N}{16(2J_\perp + J_\perp^a)} \ \frac{2- \epsilon/2
+ (\epsilon+1)\cos 2\varphi +\epsilon\cos 4 \varphi}{1- \epsilon^2/4 + \epsilon \cos 2\varphi + (\epsilon^2/2)
\cos 4\varphi}\,.
\tag{S14}\label{DEZ110}
\end{equation}
Still, a simple analysis shows that the field contribution is finite for all $\epsilon$ and selects states with  $\varphi = 0,\pi$ (see also below).
The amplitude of the $2\varphi$ harmonic exactly equals $-1/2$ the corresponding result for
$\mathbf{H} \parallel [001]$.

Finally, the quadratic energy correction for $\mathbf{H}\parallel [111]$ is
\begin{equation}
\Delta E^{[111]} = \frac{(g_\perp \mu_B H)^2 N}{8(2J_\perp + J_\perp^a)} \ \frac{-4+2\epsilon+\epsilon^2 + \epsilon^2 \cos 6\varphi}{1-3\epsilon^2/4 - (\epsilon^3/4)\cos 6\varphi} \,.
\tag{S15}\label{DEZ111}
\end{equation}
As before, the denominator does not vanish, ensuring a finite energy shift, and the selection term is proportional to $\cos 6\varphi$ with a positive coefficient.
The obtained analytic expressions for all three field directions were checked against direct numerical minimization
of the classical energy (\ref{Hmzh}) in the four-sublattice basis and full agreement was found for weak magnetic fields $H\ll H_s$.
In addition, the numerical minimization confirms a
change of sign of the $4\varphi$ harmonic  in the expression (\ref{DEZ110}) for negative $\epsilon$.
While the prefactor  of $\cos2\varphi$ is always negative, the coefficient of $\cos4\varphi$
 becomes positive for $J^\perp_a/J_\perp > 4$ ($\epsilon<0$).
The sign change can modify stability of the two domains with $\varphi = 0, \pi$ predicted by the symmetry analysis.
Indeed, more complex spin configurations were found in our simulations for $J_\perp^a/J_\perp \gtrsim 10$.
Though this case is far beyond the parameter range expected for $\mathrm{Er_2Ti_2O_7}$,
it still might be relevant for another pyrochlore material, $\mathrm{Er_2Ge_2O_7}$, which was suggested
to order in  the $\psi_3$ state \cite{Dun15s}.

\section{III.\ Field-induced orientational transitions}

Using the third-order real-space perturbation theory
we obtained the following expression for the
$Z_6$ anisotropy \cite{Maryasin14s}:
\begin{equation}
E_6(\varphi)  =  - \frac{2J_\perp+2J_\perp^a}{10^3}\, \epsilon^3\,(1+\tfrac{1}{4}\epsilon^2) \cos 6\varphi\,.
\tag{S16}\label{EQ6}
\end{equation}
Besides spin-flip hopping processes, this expression  takes into account the effect of spin-flip
interaction. Interactions reduce by $\sim 40$~\%  the amplitude of the six-fold harmonics
in comparison with the harmonic spin-wave result. We consider the above expression to be
more accurate, though the approximate nature of all analytic expressions for the quantum anisotropy
must be kept in mind.

In order to determine the transition  fields for ${\bf H}\parallel [001]$ and $[110]$
we use complete expressions  for the field-induced anisotropy terms (\ref{DEZ001}) and (\ref{DEZ110})
together with the quantum contribution (\ref{EQ6}).
By checking the stability of the high-field state with $\varphi=\pi/2$ we obtain:
\begin{equation}
g_\perp\mu_B H_c^{[001]} = \frac{3}{5} (2J_\perp+J_\perp^a)
\epsilon (1+\epsilon)\sqrt{\frac{\epsilon(1+\tfrac{1}{4}\epsilon^2)}{5(1-\epsilon)}\,}\,.
\tag{S17}
\end{equation}
It applies for $\epsilon>0$ or $J_\perp^a/J_\perp<4$.
Numerical results obtained with this expression are shown in top-left panel of Fig.~2 of the main text.

A similar calculation for the stability of the state with $\varphi=\pi$ yields
\begin{equation}
g_\perp\mu_B H_c^{[110]} = \frac{6}{5} (2J_\perp+J_\perp^a)
|\epsilon| (1+\epsilon/2)
\sqrt{\frac{|\epsilon|(1+\tfrac{1}{4}\epsilon^2)(1+\tfrac{1}{2}\epsilon)}{10(1-\epsilon)(1+\tfrac{7}{2}\epsilon)}\,}\,.
\tag{S18}
\end{equation}
This expression holds for $-2/7\leq \epsilon\leq 0$ or $4\leq J_\perp^a/J_\perp \leq 12$, the corresponding
curve is drawn of the top-right panel of Fig.~2 of the main text.
For $\epsilon\to-2/7$, the above expression diverges, which means that $H_c^{[110]}$ approaches the saturation field
$H_s$. Beyond this range for $\epsilon<-\tfrac{2}{7}$, the $\psi_2$ states do not appear in a magnetic field
and the low-symmetry states occupy the whole range of magnetic fields $0<H<H_s$.

Finally, for ${\bf H}\parallel [111]$, in addition to the contribution (\ref{DEZ111}), an external magnetic field also generates a
three-fold harmonic  in the angular-dependant part of energy (see the main text):
\begin{equation}
E_3(\varphi) = \lambda_3 \frac{(g_\perp \mu_B H)^3}{(2J_\perp + J_\perp^a)^2}\, \cos 3\varphi \,.
\tag{S19}\label{E33}
\end{equation}
The dimensionless factor $\lambda_3$  cannot be expressed analytically. Therefore, we proceed
with calculations in the two-step manner.
First, we consider only $E_3(\varphi)$ and (\ref{DEZ111}) contributions.
Their competition produces a single transition exhibited by the classical model at zero-temperature,
{\it i.e.}, in the absence of the zero-field six-fold anisotropy:
\begin{equation}
g_\perp\mu_B H_c^{[111]} = (2J_\perp+J_\perp^a) \frac{2\epsilon^2(2-\epsilon)}
{\lambda_3(1-\epsilon)(2+\epsilon)^3} \,.
\tag{S20}
\end{equation}
Numerical differentiation of the calculated magnetization curves is used to determine the position of this transition
for all values of $J_\perp^a/J_\perp$, an example is given in the inset of Fig.~1.
This allows us to estimate the dimensionless amplitude $\lambda_3$ of the three-fold harmonic as a function of $J_\perp^a/J_\perp$.
Second, we add the zero-field term (\ref{EQ6}) and use
the numerical values of $\lambda_3$ to solve the cubic equation obtained by determining the stability
of the state with $\varphi=\pi$. Depending on the coefficients, there are either no or two positive real roots of that equation for
$\epsilon>0$, whereas for $\epsilon<0$ the cubic equation has always one positive root.
The obtained numerical results are used for Fig.~2 of the main text.

\section{IV.\ Classical Monte Carlo simulations}

\begin{figure}[t]
\centerline{
\includegraphics[width=0.35\columnwidth]{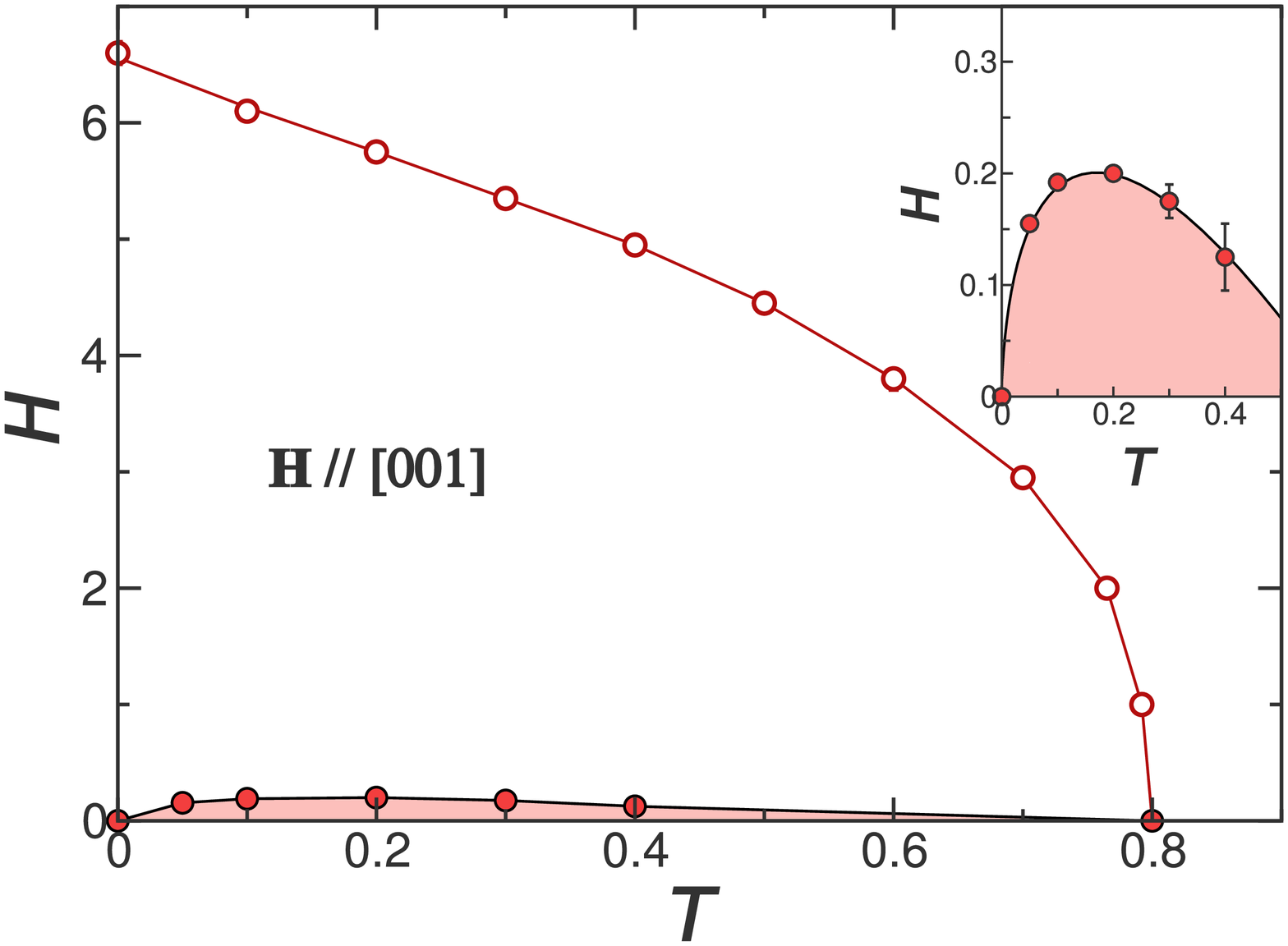} \hspace{15mm}
\includegraphics[width=0.35\columnwidth]{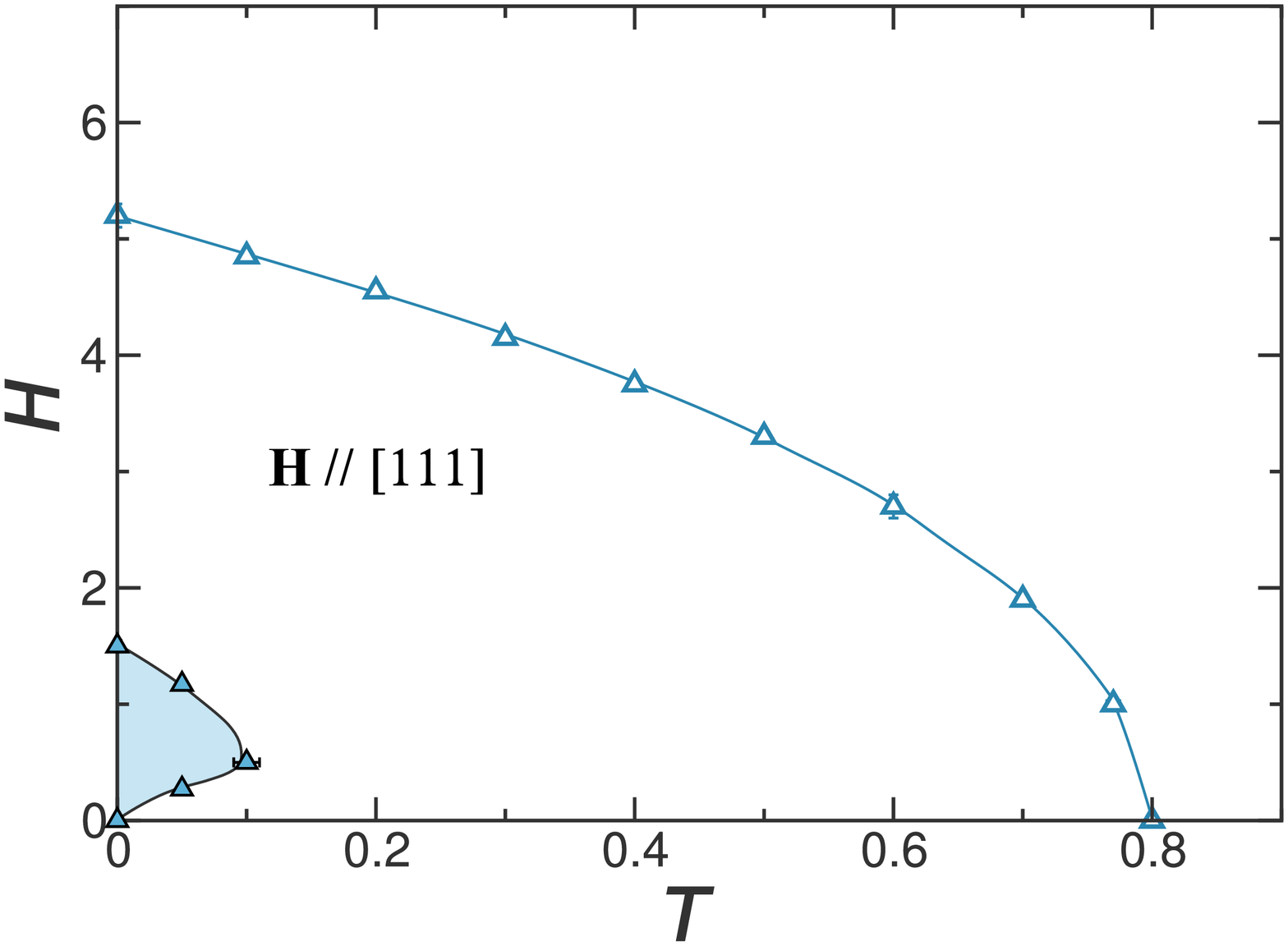}
}
\caption{Monte Carlo phase diagrams of the $XY$ pyrochlore antiferromagnet
for two orientations of an applied magnetic field.}
\label{fig:PD}
\end{figure}

Monte Carlo simulations of the classical model (\ref{Hmzh}) were performed using
a hybrid algorithm, which consists of a combination of single spin Monte Carlo updates
with microcanonical overrelaxation steps \cite{Zhitomirsky12s, Maryasin14s,Zhitomirsky14s}.
Periodic clusters with $N = 4L^3$ spins up to $L = 24$ were simulated.
Majority of Monte Carlo runs were performed at fixed $T$ starting with a random initial spin configuration
at a large enough field $H > H_s$, and decreasing gradually the field strength. Statistical averages
were taken over 300 independent runs, each measurement was taken during up to $2\cdot 10^5$ Monte Carlo steps
and the first $5\cdot 10^4$ steps at each field were omitted for thermalization.

The transition from the paramagnetic to the ordered phase was determined with the help of
the $\Gamma_5$-representation order parameter
\begin{equation}
m = \sqrt{m_1^2 + m_2^2}, \qquad \qquad m_1 = \frac{1}{N} \sum_i \mathbf{S}_i \cdot \mathbf{\hat{x}}_i, \qquad  m_2 = \frac{1}{N} \sum_i \mathbf{S}_i \cdot \mathbf{\hat{y}}_i.
\tag{S21}\label{HP.m}
\end{equation}
Different antiferromagnetic ordered phases were distinguished by simultaneously measuring
the clock-type order parameters $m_p^\prime$, $m_p^{\prime \prime}$
\begin{equation}
m_p^{\prime} = \frac{1}{m^{p-1}} {\rm Re} \lbrace(m_1 + im_2)^p\rbrace = m\cos p\varphi; \qquad \qquad
m_p^{\prime\prime} = \frac{1}{m^{p-1}}{\rm Im} \lbrace(m_1 + im_2)^p\rbrace  =m \sin p\varphi
\tag{S22}\label{HP.M''}
\end{equation}
with $p = 2, 3, 6$. Finally, the corresponding Binder cumulants $U(m_p) = 1- \langle m_p^4 \rangle / 3\langle m_p^2 \rangle^2$ were
used for precise determination of the phase boundaries.

\begin{figure}[b]
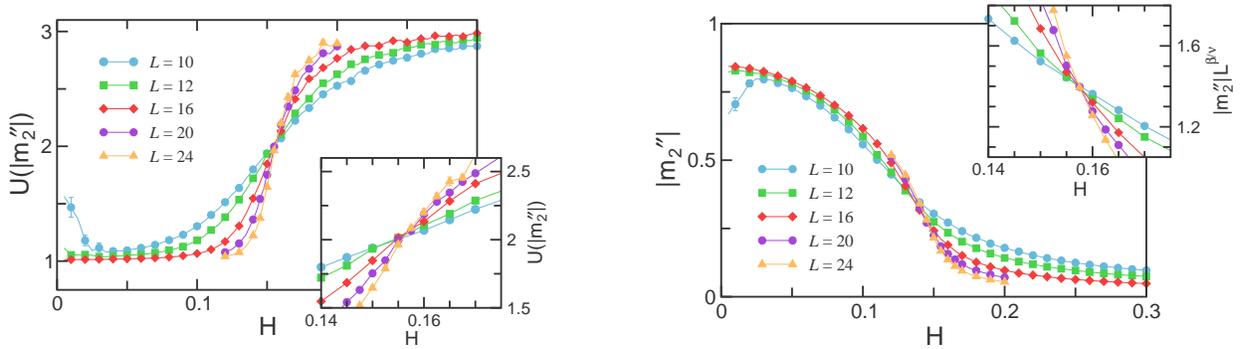

\centerline{
\includegraphics[width=0.4\columnwidth]{MC001_Ub} \hspace{12mm}
\includegraphics[width=0.43\columnwidth]{MC001_M2}
}
\caption{Finite-size analysis of the orientational transition for $\mathbf{H} \parallel [001]$ at $T = 0.05$.
Left panel: field dependence of the Binder cumulants for the  $m_2^{\prime\prime}$ order parameter
for different cluster sizes. Their crossing defines the transition field $H_c = 0.156(2)$.
Right panel: field dependence of the  order parameter $|m_2^{\prime\prime}|$.
The inset illustrates the procedure for determining the $\beta/\nu$ value.
The best crossing is obtained for $\beta/\nu \approx 0.7$.
}
\label{fig:MCC}
\end{figure}

The spin Hamiltonian parameters adopted in the simulations are $J_\perp = 1$, $J_\perp^a = 1.5$,
$g_\perp = 1$, $g_z = 0.5$ in accordance with the magnetization fits (Sec.~I). Complete phase
diagrams for the two field orientations $\mathbf{H} \parallel [001]$ and $\mathbf{H} \parallel [111]$
are shown in Fig.~\ref{fig:PD}. The zero-field transition temperature is $T_c=0.793(2)$ for this
set of exchange parameters \cite{Zhitomirsky14s}. Nontrivial phases are denoted by color shading.
Phase transitions are second order except for the PM-AFM phase boundary for $\mathbf{H} \parallel [111]$.
The latter boundary is a line of first-order phase transitions with a small discontinuous jump in the magnetization, although
the discontinuity is barely observable for $T \gtrsim 0.2$.

Figure \ref{fig:MCC} illustrates the finite-size analysis used to determine the boundary $H_c(T)$
of the orientational transition for ${\bf H}\parallel [001]$. Temperature is set to
$T = 0.05$. The transition field $H_c = 0.156(2)$ is determined from the crossing of the Binder cumulants
for different cluster sizes (left panel).
The right panel shows the field dependence of the order parameter $|m_2^{\prime\prime}|$ and the inset illustrates
the scaling procedure used to obtain an estimate for the  critical exponent ratio  $\beta/\nu = 0.70(5)$.
This value is obtained by searching for the best crossing of the scaled order parameters $|m_2^{\prime\prime}| L^{\beta/\nu}$
at the same critical field $H_c$ upon varying $\beta/\nu$. It differs  from the value $\beta/\nu = 0.518$ for
the Ising universality class in three dimensions. The origin for such a substantial discrepancy is not clear at present.
Most probably it is due to the fact that the six-fold anisotropy is dangerously irrelevant in 3D, see, {\it e.g.},
\cite{Zhitomirsky14s} and thus the correct scaling behavior is only obtained for significantly larger
lattices.

\end{document}